\documentclass[sigconf, review=false]{acmart}
\usepackage{graphicx} \usepackage{xcolor}

\title{Hypothesis-Driven Shelf Generation for Personalised Recommendation}
\copyrightyear{2026}
\acmYear{2026}
\setcopyright{cc}
\setcctype{by-nc-nd}
\acmConference[RecSys '26]{20th ACM Conference on Recommender Systems}{September 27-October 02, 2026}{Minneapolis, MN, USA}
\acmBooktitle{20th ACM Conference on Recommender Systems (RecSys '26), September 27-October 02, 2026, Minneapolis, MN, USA}
\acmDOI{10.1145/3773078.3831914}
\acmISBN{979-8-4007-2284-4/2026/09}

\author{Aleksandr V. Petrov, Tarun Chillara, Matthew D. Moellman, Lucas de Haas, Yabai Song, Alina Susoykina, Melissa Crawford, Gabriel Negash, Erik Franco, Tasnim Rahman, Binal Jhaveri, Shubham Bansal, Hugues Bouchard, Roberto Mirizzi, Mounia Lalmas, Aloïs Gruson}
\affiliation{
    \institution{Spotify}
\country{UK, US, Netherlands, Spain, France}
}
\email{aleksandrv,tarunc,mmoellman,lucasd,yabais,alinas,melissac,gnegash,}
\email{efranco,tasnimr,binalj,shubhamb,hb,rmirizzi,mounial,agruson@spotify.com}

\begin{abstract}

  Modern recommendation interfaces organise content into shelves: themed rows such as ``More of What You Like'' or ``New Releases for You.'' In production systems, these shelves are typically defined through hand-crafted templates coupled with dedicated retrieval logic. While effective for broad recommendation intents, this approach does not scale to the long tail of individual taste.

  We present a \emph{content-hypothesis-driven} shelf generation system for Spotify Home that replaces fixed templates with natural-language hypotheses describing what a personalised shelf should contain. The system has four stages---hypothesis generation, catalogue fulfilment, shelf alignment, and offline serving. This decomposition decouples shelf planning from catalogue fulfilment, supports independent optimisation of planning and retrieval, and enables both constrained generative retrieval over catalogue entities and distillation of frontier LLM behaviour into compact models.

Our production pipeline combines hypothesis generation, generative retrieval, candidate selection and shelf alignment, offline LLM-as-a-judge evaluation, and precomputed serving. We describe the end-to-end architecture and evaluate it through offline analyses and an early online evaluation under uniform random exposure on Spotify Home. Results show that hypothesis-driven shelves substantially expand personalised recommendation supply with engagement that varies by content type and is competitive with strong existing shelves in some settings.
\end{abstract}

\begin{document}

\maketitle

\begin{figure}[H]
    \centering
    \includegraphics[width=0.54\linewidth]{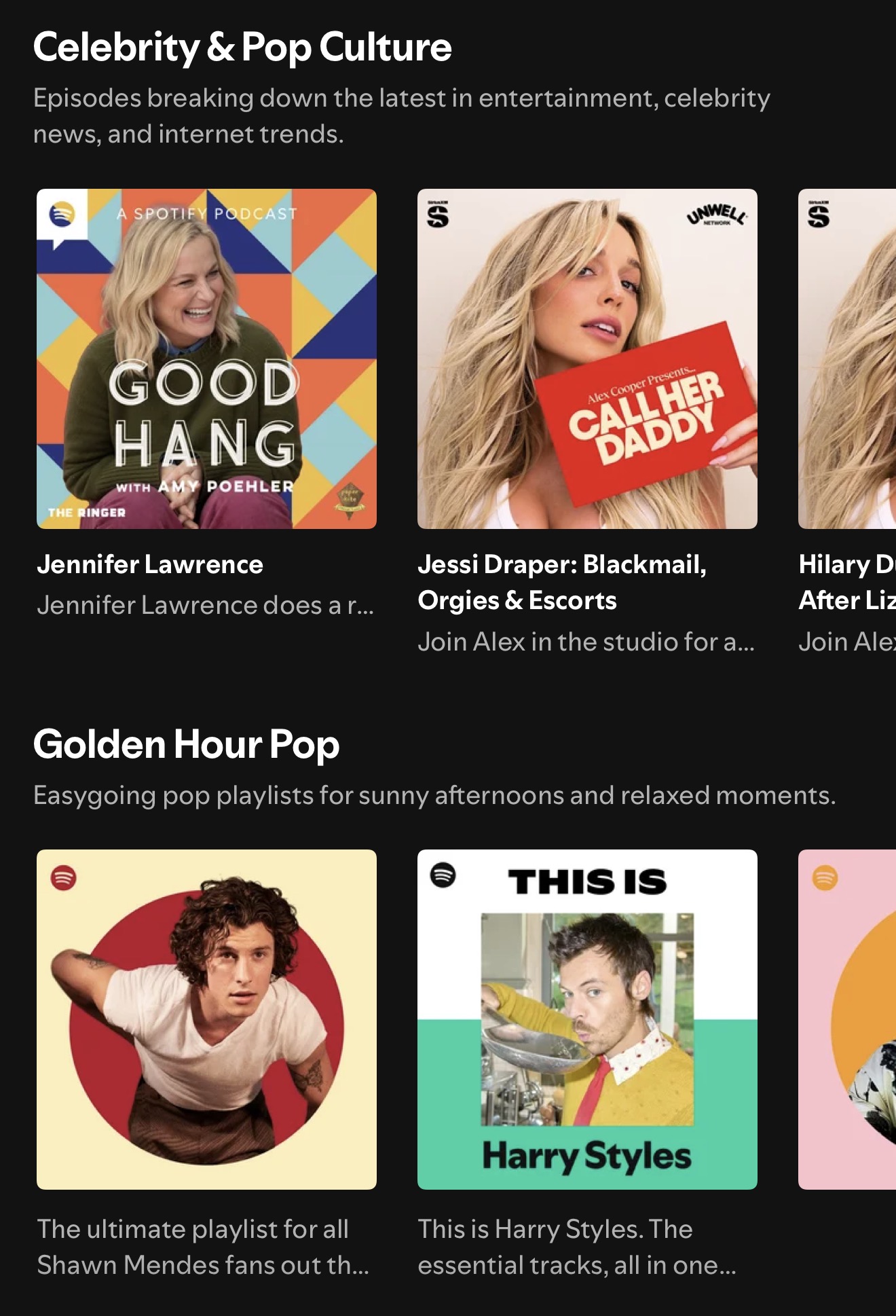}
    \caption{Examples of hypothesis-driven personalised shelves in Spotify Home}
    \Description{Spotify Home screenshot showing two hypothesis-driven shelves: podcast episodes about celebrity and pop culture, and easygoing pop playlists.} \label{fig:shelf-examples} \end{figure} 
\section{Introduction}

Recommendation surfaces increasingly rely on \emph{shelves}: horizontal rows that offer compact, themed entry points into a large catalogue. A single homepage may combine shelves for familiar favourites, new releases, editorial playlists, podcast episodes, and long-tail discovery. The shelf format is powerful because it makes recommendation intent legible. A row title such as ``More of What You Like'' or ``New Releases for You'' is not merely a user-interface label; it is a promise about why the items belong together and on this user’s page.

In most production systems, however, this promise is implemented through a finite set of hand-designed templates. Each template implicitly couples three decisions that are often treated as inseparable: what user need the shelf should satisfy, what content should count as a valid realisation of that need, and what retrieval or ranking system should populate the row. While effective for broad and recurring intents such as recent listening or globally popular releases, this design creates a bottleneck for the long tail of individual taste. Maintaining dedicated templates for every combination of genre, era, scene, mood, familiarity, market, and media type is infeasible at production scale.

To overcome these limitations, we instead formulate shelf generation as a generative planning problem. Rather than selecting from a fixed set of shelf templates, a model generates content hypotheses: natural-language descriptions of personalised shelf concepts. These hypotheses may describe broad themes, such as contemporary dance-pop playlists, or highly specific niches, such as glacial ambient post-rock albums with orchestral textures. Figure~\ref{fig:shelf-examples} shows examples on Spotify Home.

Each hypothesis acts as an intermediate planning representation between user modelling and catalogue retrieval. The challenge, however, is that a useful shelf is not merely a plausible textual description. It must be grounded in real catalogue entities, satisfy product and user constraints, produce faithful user-facing text, scale to millions of users, and integrate into a production system without introducing additional serving latency.

These requirements motivate the architectural decomposition of our system: separating personalised shelf planning from catalogue fulfilment. A hypothesis generator uses signals such as listening history, recent affinities, podcast behaviour, and market information to produce shelf hypotheses containing user-facing text together with retrieval constraints and metadata. A separate fulfilment model then retrieves catalogue entities that realise the generated hypothesis through constrained generative retrieval.

This separation is central to the architecture. The hypothesis generator can focus on producing personalised shelf concepts, while the fulfilment stage independently optimises for accurate and scalable retrieval over large and heterogeneous catalogues.

The production pipeline runs fully offline. For each user, the system generates shelf hypotheses, retrieves matching catalogue items, optionally refines the resulting shelves through an LLM-based alignment step, and precomputes the final shelves for serving. Performing these stages offline makes it possible to use broader retrieval and post-processing without introducing additional latency when users open the Spotify app. The generated shelves are introduced into Home as additional ranking candidates rather than fixed placements, allowing them to compete with existing shelves through Spotify’s ranking and presentation systems. This allows the pipeline to expand the set of personalised shelves considered by Home while avoiding online LLM inference during serving.

Moving from prototype LLM recommendations to production-quality shelves introduces several practical challenges. The system must ensure that retrieved items remain consistent with the intended shelf type, respect familiarity and freshness constraints, and maintain alignment between shelf titles and the items ultimately shown to users.
Our implementation addresses these challenges through a combination of constrained retrieval, content-type-specific filtering, and a downstream alignment stage that selects the final shelf items and rewrites user-facing text to better reflect the retrieved items.

We make the following contributions:
\begin{itemize}
  \item We formulate shelf generation as a hypothesis-driven recommendation problem that separates personalised shelf planning from catalogue retrieval.
  \item We introduce an architecture that combines hypothesis generation, constrained generative retrieval, and LLM-based shelf alignment across music and podcast types.
  \item We evaluate the system through offline analyses and an early online evaluation under uniform random exposure on Spotify Home, showing that hypothesis-driven shelves expand personalised recommendation supply while remaining compatible with production serving, latency, and quality constraints.
\end{itemize}

More broadly, our work positions shelf generation at the intersection of multi-list recommendation, language-based recommendation, and generative retrieval, treating shelf hypotheses as intermediate planning representations within a production recommendation system.

\section{Related work}
\label{sec:related-work}

This section situates our work within the broader literature on multi-list recommendation, language-based recommendation, generative retrieval, and industrial recommender systems.\\

Prior work on multi-list recommendation studies recommender interfaces organised as pages of labelled rows, often referred to as 
\emph{carousels} or \emph{shelves}, rather than as a single ranked list.
Existing research formalises the benefits of multi-list interfaces~\citep{rahdari2022magic}, 
surveys their design space~\citep{loepp2023multilist}, 
and analyses how users interact with carousel layouts~\citep{loepp2023carousel}. 
In music recommendation, carousel personalisation has been formulated as a contextual bandit problem~\cite{bendada2020carousel}, 
while automatically generated collections have been shown to outperform item-level recommendation~ \cite{singal2021collection}. 
Unlike these approaches, our system does not simply rank or personalise a fixed set of shelves. Instead, it generates personalised shelf hypotheses and subsequently fulfils them with catalogue items.

A second line of work treats recommendation as a language problem. Recommendation tasks have been formulated within a unified text-to-text frameworks, showing that natural language can serve as a general interface for recommendation~\cite{geng2022p5}. 
In music recommendation, recent work has explored retrieval from free-form prompts, including Text2Tracks~\citep{palumbo2025text2tracks}, 
Text2Playlist~\citep{delcluze2025text2playlist}, and language-model-based playlist generation from playlist titles~\citep{charolois2025playlistlm}. 
These approaches typically begin from user-provided language and directly retrieve tracks or playlists. In contrast, our system generates shelf hypotheses automatically from user profiles and uses them as intermediate planning representations for shelf creation.

Our fulfilment stage is most closely related to generative retrieval. Prior work on neural corpus indexing and differentiable retrieval shows that sequence models can generate identifiers directly rather than retrieve items through conventional indexes~\citep{tay2022dsi,wang2022nci}. 
In recommendation, Semantic-ID-based generative retrieval has been introduced as an alternative to traditional retrieval architectures~\cite{rajput2023generativeretrieval}, with subsequent work showing that Semantic IDs improve recommendation 
generalisation~\citep{singh2024semanticids,d2026deploying} and support joint search-and-recommendation settings~\citep{penha2025jointsemanticids}.

We build on this line of work by treating shelf fulfilment as constrained generation over catalogue identifiers. However, generative retrieval plays a different role in our system: fulfilment is conditioned only on the generated shelf hypothesis rather than on the full user profile, deliberately separating personalised shelf planning from catalogue retrieval.

Our work is also related to slate generation and industrial multi-stage recommender architectures. Recent work on slate generation produces coherent item sets directly from natural-language prompts~\cite{tomasi2025prompttoslate}, whereas our system first generates a personalised shelf concept and then retrieves items that instantiate it. More broadly, industrial recommender systems such as Netflix~ \citep{gomezuribe2016netflix} and
YouTube~\citep{covington2016youtube} demonstrate the value of decomposing recommendation into multiple stages. Our architecture follows a similar principle, but with a decomposition centred on personalised shelf planning, hypothesis-driven retrieval, and post-hoc shelf alignment. The objective is not only to recommend relevant items, but also to generate shelf concepts that are coherent, interpretable, and meaningful within a Home interface.

\section{System Architecture}
  \label{sec:pipeline}

  This section describes the architecture of the proposed shelf generation system. The pipeline is organised around an explicit decomposition between personalised shelf planning, catalogue fulfilment, shelf alignment, and serving. Figure~\ref{fig:pipeline-overview} summarises the resulting production flow, while Table~\ref{tab:pipeline-eval-map} formalises the responsibilities, interfaces, and evaluation boundaries associated with each stage. The remainder of the section describes these stages in sequence and explains how the decomposition supports the stage-specific evaluation methodology developed in Section~\ref{sec:results}.

  \subsection{Design goals and decomposition}

To support hypothesis-driven shelf generation, we organise the system around an explicit intermediate representation: the \emph{shelf hypothesis}. 
  Section~\ref{sec:related-work} identifies four requirements not fully addressed by prior work: generating new shelf concepts rather than selecting from a fixed inventory, inferring shelf concepts from behavioural profiles rather than user-written prompts, grounding those concepts in a large catalogue through constrained retrieval, and supporting production-scale serving together with stage-specific evaluation. These requirements motivate the central architectural decomposition of the system: \emph{personalised shelf planning} followed by \emph{hypothesis-driven catalogue fulfilment}.
  
Under this decomposition, the planning stage determines what shelves should exist for a user, while the fulfilment stage retrieves catalogue entities that instantiate those shelf concepts. The shelf hypothesis forms the interface between these stages: a representation combining free-form natural-language intent with the metadata required for retrieval, filtering, routing, and evaluation.

  For a user \(u\), the planner constructs a taste profile \(p_u\) from signals such as 
  recent listening, long-term affinities, market context, familiar content, and podcast engagement. Using this profile, the planner generates shelf hypotheses containing a natural-language shelf description $q$, a target content type $c$, a familiarity level $f$, optional routing constraints $r$ such as market or freshness, and provisional titles and subtitles $t_0, d_0$. Formally, we represent each hypothesis as
  \[
    h = (q, c, f, r, t_0, d_0),
  \]
The hypothesis is therefore not merely a retrieval query: it is a compact contract between planning and fulfilment. It must be expressive enough to capture long-tail user taste, specific enough to constrain retrieval, and structured enough to support evaluation.

  Fulfilment maps \(h\) to a candidate set of catalogue entities, and candidate selection and shelf alignment convert that set into a final shelf record
  \[
    s = (t, d, q, \mathcal{I}),
  \]
  where \(t,d\) are the final title and subtitle and
  \(\mathcal{I}\) is an ordered list of resolved Spotify catalogue entities.
  This separation lets the planner optimise for taste alignment, specificity, and
  discovery, while fulfilment optimises for catalogue validity, relevance, and
  coverage.

\begin{figure*}[t]
  \centering
  \includegraphics[width=1.0\textwidth]{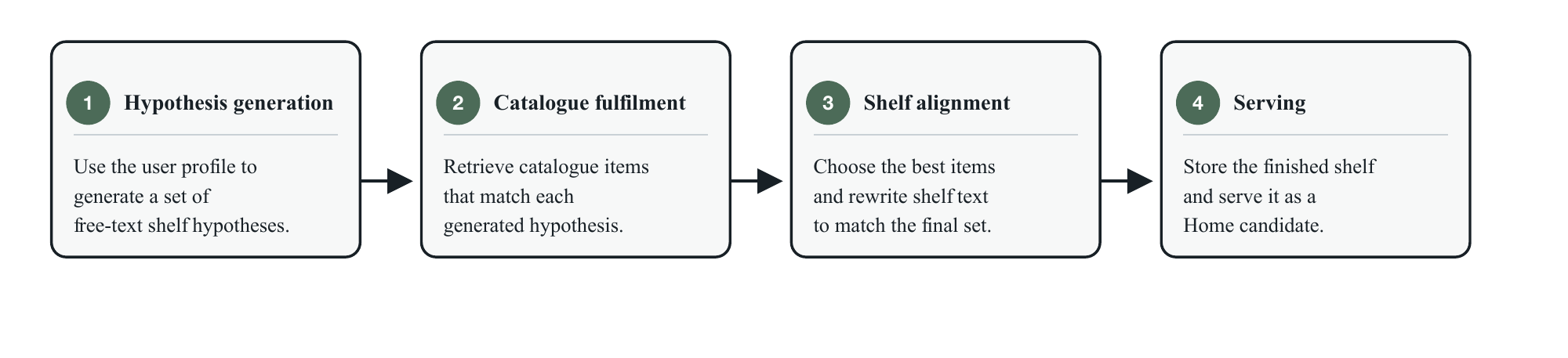}
  \caption{Conceptual pipeline. User-profile information is consumed only by
  hypothesis generation. Downstream fulfilment, alignment, and serving operate
  on shelf hypotheses and candidate catalogue entities.}
  \Description{Pipeline diagram in which a user taste profile feeds hypothesis generation, followed by catalogue fulfilment, candidate selection and shelf alignment, and precomputed serving on Home.} \label{fig:pipeline-overview}
  \end{figure*}

  Figure~\ref{fig:pipeline-overview} summarises the production flow. The system operates as a daily batch pipeline over eligible users, allowing larger retrieval candidate sets, metadata enrichment, post-processing, and offline analysis without introducing latency during Home serving. Each stage produces a stable intermediate artefact (i.e., hypotheses, retrieved candidate sets, aligned shelves, and serving records) allowing downstream components and evaluation procedures to evolve independently. 
  Table~\ref{tab:pipeline-eval-map} formalises the resulting stage boundaries and previews the analyses presented in Section~\ref{sec:results}.

The following subsections describe each stage in detail, from personalised hypothesis generation through serving on Home.

  \begin{table*}[t]
  \centering
  \caption{Pipeline responsibilities and the evaluation boundaries used in
  Section~\ref{sec:results}.}
  \label{tab:pipeline-eval-map}
  \scriptsize
  \setlength{\tabcolsep}{2.5pt}
  \renewcommand{\arraystretch}{1.08}
  \begin{tabular}{@{}p{0.085\textwidth}p{0.135\textwidth}p{0.15\textwidth}p{0.205\textwidth}p{0.205\textwidth}p{0.165\textwidth}@{}}
  \toprule
  Stage & Input & Output & Responsibility & Main failure mode & Evaluation hook \\
  \midrule
  1. Hypothesis &
  User profile &
  Shelf hypotheses &
  Infer personalised shelf intents from behavioural evidence &
  Generic, unsupported, or underspecified concepts &
  User-to-Hypothesis Judge \\
  \midrule
  2. Fulfilment &
  Generated hypothesis &
  Candidate URIs &
  Retrieve catalogue entities that instantiate the proposed concept &
  Items fail to realise the hypothesis &
  Hypothesis-to-Shelf Judge; retrieval baselines \\
  \midrule
  3. Shelf alignment &
  Resolved candidate URIs and draft shelf text &
  Final items and revised shelf text &
  Select a coherent shelf and align the visible promise with the items &
  The title overclaims or the row lacks set-level coherence &
  Pre/post alignment comparison \\
  \midrule
  4. Serving &
  Final shelf record &
  Home candidate &
  Inject precomputed shelves into Home ranking &
  Strong offline shelves may not translate to competitive engagement &
  Uniform random exploration \\
  \bottomrule
  \end{tabular}
  \end{table*}

  \subsection{Stage 1: Hypothesis generation}

The first stage generates a small set of shelf hypotheses from a user profile \(p_u\). 
Generating multiple hypotheses allows the system to cover different aspects of a user’s taste profile within a single Home surface. Each hypothesis specifies a target content type, familiarity level, optional freshness constraints, provisional title \(t_0\), provisional subtitle \(d_0\),
  and natural-language shelf hypothesis \(q\). For example, a user with strong affinity for Nordic ambient, post-rock, and modern classical music may receive a hypothesis such as ``glacial ambient post-rock with orchestral textures.'' Unlike a fixed shelf template, the hypothesis can express narrow genre intersections, artist neighbourhoods, eras, languages, moods, and listening contexts.

Early versions of the system used a frontier LLM in batch mode for this stage. To support production-scale generation, we distil this behaviour into a compact open-source LLM deployed on a GPU batch cluster. The distilled model receives a compact serialisation of the user profile and emits structured shelf hypotheses. The schema constrains the generator to the fields required by downstream stages while still allowing free-form hypothesis text. Invalid or missing categorical values are mapped to safe defaults, and generated hypotheses are grouped into per-user bundles for downstream retrieval and filtering.

  \subsection{Stage 2: Catalogue fulfilment}

The fulfilment stage operationalises the shelf hypothesis by grounding it in concrete catalogue entities. Given a generated shelf hypothesis, the objective of fulfilment is to retrieve items that instantiate the proposed shelf concept while respecting content-type, familiarity, and routing constraints.

We formulate fulfilment as a generative retrieval problem: rather than scoring all candidate items directly, the model generates Semantic IDs (SemIDs), compact discrete identifiers learned over catalogue entities. Retrieval is therefore performed through sequence generation over SemIDs rather than direct scoring over the full catalogue, after which generated identifiers are resolved into Spotify catalogue items.

The fulfilment model is built on a smaller open-source LLM whose vocabulary is extended with SemIDs grounded in Spotify catalogue data using a procedure similar to~\cite{he2026plum,d2026deploying}. To ensure that generated identifiers always correspond to valid catalogue entities, decoding is constrained using content-type-specific indexes implemented as tries. Separate indexes are maintained for albums, artists, editorial playlists, podcast shows, and podcast episodes.

The fulfilment model receives the shelf hypothesis together with metadata describing the target shelf type and retrieval constraints, but not the full user profile. This separation is intentional: user-specific reasoning has already occurred during hypothesis generation. Fulfilment therefore solves a narrower problem: retrieving catalogue entities that realise the proposed shelf concept.

Different shelf types are fulfilled against different constrained indexes. Editorial shelves use market-specific editorial indexes, fresh-release shelves use recency-filtered indexes, familiar shelves decode against on-the-fly tries constructed from user-familiar catalogue entities, and discovery shelves use broader catalogue indexes. Generated SemIDs are then resolved to Spotify URIs and filtered for type consistency, prior listening, and familiarity constraints.

Together, these constraints ensure that fulfilment remains faithful to the generated shelf hypothesis while producing valid, type-consistent, and production-ready catalogue recommendations.

\subsection{Stage 3: Candidate selection and shelf alignment}

  Constrained fulfilment produces a ranked candidate set rather than a final display list. While many retrieved items may individually match the shelf hypothesis, the resulting shelf can still fail at the level of the complete recommendation row. In particular, the displayed shelf title may overstate, understate, or otherwise mischaracterise the retrieved item set.

To address this problem, the candidate selection and shelf alignment stage jointly optimises the final shelf contents and user-facing text. Given the original hypothesis, provisional title \(t_0\), provisional subtitle \(d_0\), and enriched candidate metadata, an LLM selects the final \(k\) items and rewrites the title and subtitle to produce \(t,d\). 

This stage addresses a challenge specific to shelf recommendation rather than generic ranked retrieval. A shelf title makes a promise about the entire row: even when retrieved items are reasonable, the shelf can still fail if its displayed text describes a narrower, broader, or different concept than the resolved item set. Candidate selection and shelf alignment therefore optimise the shelf as a coherent recommendation unit, balancing item relevance, set coherence, diversity, and title-promise fulfilment.

The alignment stage creates a cleaner separation between retrieval and presentation. Fulfilment is responsible for retrieving entities that realise the shelf concept, while alignment ensures that the shelf text faithfully reflects the final retrieved set. This separation allows retrieval quality and presentation quality to be analysed independently in the evaluation framework introduced in Section~\ref{sec:results}.

  \subsection{Stage 4: Serving}

The final stage integrates generated shelves into the Home recommendation surface. Because shelf planning, fulfilment, and alignment are completed offline, serving only needs to retrieve precomputed shelf candidates for ranking and presentation on Home. The shelves are not pinned placements; instead, they compete with other Home candidates through the platform’s ranking mechanisms. This design makes deployment incremental: the system can expand the supply of personalised shelf candidates without introducing additional latency during Home serving.

The serving stage also preserves the separation between offline shelf generation and online ranking. Shelf engagement is strongly influenced by shelf position and by how the production ranking system selects and orders shelves, making direct comparison between shelf families difficult. To reduce these effects, the online evaluation in Section~\ref{sec:results} uses uniform random exploration, where shelf ordering is randomised independently of the production ranker. This allows shelf performance to be compared under a known exposure policy.

The same structured shelf representation is maintained across fulfilled shelves, aligned shelves, and serving records. This modularity allows individual stages to be swapped or ablated (e.g., frontier versus distilled hypothesis generation, constrained versus unconstrained fulfilment, or with versus without shelf alignment) without changing downstream serving behaviour. Table~\ref{tab:pipeline-eval-map} summarises the resulting evaluation boundaries and their corresponding analyses.

Taken together, the four stages transform behavioural evidence into shelf hypotheses, ground those hypotheses in catalogue entities, refine them into coherent shelves, and expose the candidates to Home ranking. Section~\ref{sec:results} evaluates each stage of this decomposition, from hypothesis quality to online shelf performance.

\section{Evaluation and Results}
\label{sec:results}

The architectural decomposition introduced in Section~\ref{sec:pipeline} induces a corresponding decomposition of the evaluation problem. Each stage of the pipeline produces a distinct intermediate artefact and failure mode, allowing hypothesis generation, catalogue fulfilment, shelf alignment, and online serving to be evaluated independently. Table~\ref{tab:pipeline-eval-map} summarises these evaluation boundaries and their associated methodologies.

We structure the evaluation around four questions:
\begin{enumerate}
    \item Does the hypothesis generator produce personalised and sufficiently specific shelf concepts? \item Does Generative Retrieval fulfil those hypotheses more effectively than lexical and embedding-based retrieval baselines?
    \item Does candidate selection and shelf alignment improve the coherence and presentation quality of the final shelf?
    \item Do the resulting shelves perform competitively on Home under uniform random exposure?
\end{enumerate}

To answer these questions, we combine two evaluation methodologies. Offline LLM-as-a-judge analysis evaluates hypothesis quality, catalogue fulfilment, and shelf alignment at the corresponding pipeline stages. Online behavioural analysis under uniform random exploration evaluates how shelves perform under uniform random exposure on Home. The remainder of the section follows this decomposition, progressing from offline evaluation of intermediate artefacts to online evaluation under the same exposure policy.
 
\subsection{LLM-as-a-Judge Evaluation Methodology}
\label{sec:llm-judge}

We use LLM-as-a-judge evaluation because the shelf-generation task does not admit a simple gold-standard offline target. Unlike conventional recommendation settings, our system does not rank items against a fixed inventory of pre-authored shelves with historical relevance labels or canonical ``correct'' outputs. Instead, it generates new shelf hypotheses and grounds them in catalogue entities. In this open-ended setting, standard offline metrics such as recall against held-out items or ranking quality over fixed candidate sets do not directly measure whether a generated shelf concept is appropriate, whether retrieved items realise that concept, or whether the resulting shelf would make sense on a Home surface.

 Shelf quality is fundamentally set-level rather than pointwise. A shelf may fail even when several individual items are plausible: the row may be internally incoherent, omit canonical items, lack useful diversity, or fail to honour the semantic promise implied by its title and subtitle. Conversely, a shelf may succeed because its items work together as a coherent recommendation unit, a property not captured by item-level relevance scores. We therefore evaluate the pipeline using structured judge rubrics designed around the decomposition introduced in Section~\ref{sec:pipeline}.

This evaluation strategy is supported by recent work showing that LLM-generated relevance judgements can track human judgements and preserve recommender-ranking comparisons, including in industrial recommendation settings~\citep{penha2025llmjudges}. More broadly, strong LLM judges have been shown to approximate human preference judgements in open-ended generation tasks, while still exhibiting known biases that require careful protocol design~\citep{zheng2023judging}. We therefore use LLM judges as directional offline signals for optimisation and failure-mode analysis across the earlier pipeline stages, while treating online user behaviour as the final measure of shelf quality.

The decomposition introduced in Section~\ref{sec:pipeline} induces two complementary evaluation boundaries: whether the generated shelf hypothesis is appropriate for the target user, and whether the retrieved shelf successfully realises that hypothesis. We therefore develop two judges operating over the two main intermediate artefacts in the pipeline: shelf hypotheses and fulfilled shelves.

Both judges operate on a 0–2 ordinal scale ($0 = \textit{fail}$, $1 = \textit{fair}$, $2 = \textit{good}$), chosen to provide a useful balance between discriminative power and score stability. Coarser scales reduce rubric ambiguity and improve agreement between human and LLM raters, while finer-grained scales tend to amplify disagreement around intermediate categories rather than resolve it~\cite{sharma2026researchrubrics}. Restricting outputs to a small number of well-separated categories also produces more stable judge behaviour across prompt formulations and repeated runs~\cite{wang2025}. At the same time, the 0–2 scale retains sufficient resolution to distinguish failing, marginal, and acceptable outputs without introducing the calibration drift often observed at finer granularities.

\paragraph{User-to-Hypothesis Judge} This judge evaluates whether a generated shelf hypothesis is appropriate for a target user. It conditions on recent user engagement signals (e.g., top artists, top shows, recently consumed items) together with the generated shelf hypothesis and title, but does not observe retrieved shelf items, thereby isolating evaluation of the hypothesis-generation stage. Table~\ref{tab:user-to-hypothesis-rubric} summarises the five evaluated dimensions together with the overall judge score.

\begin{table}[t]
\centering
\caption{User-to-Hypothesis Judge rubric.}
\label{tab:user-to-hypothesis-rubric}
\scriptsize
\setlength{\tabcolsep}{4pt}
\renewcommand{\arraystretch}{1.08}
\begin{tabular}{@{}p{0.28\columnwidth}p{0.66\columnwidth}@{}}
\toprule
Dimension & What it measures \\
\midrule
Taste Alignment &
Whether the hypothesis is supported by direct evidence in the user's history; a
maximum score requires multiple distinct signals. \\
Personalisation Depth &
Whether the hypothesis captures a distinctive user-specific intersection of
attributes rather than a broad editorial framing. \\
Discovery Potential &
Whether the hypothesis extends the user into adjacent but plausible territory,
penalising both redundancy and untethered novelty. \\
Hypothesis Specificity &
Whether the hypothesis carries enough multi-axis content signal to constrain
retrieval without collapsing into an item-level query. \\
Title Quality &
Whether the user-facing title is engaging, representative, and comprehensible. \\
\bottomrule
\end{tabular}
\end{table}

\begin{table*}[!t]
\centering
\caption{Catalogue fulfilment quality under the Hypothesis-to-Shelf Judge (0--2 scale). Cells report mean score $\pm$ the 95\% bootstrap confidence-interval half-width from 10k resamples. Best results are bolded and second-best results are underlined. \textsuperscript{***} indicates that Generative Retrieval exceeds the strongest non-generative baseline for that metric under a two-sided paired $t$-test with Bonferroni correction ($p<0.001$ after correction).}
\label{tab:retrieval-all-dims}
\scriptsize
\resizebox{\textwidth}{!}{\begin{tabular}{lcccccccc}
\toprule
Method & Overall & Style & Relev. & Coher. & Coverag. & Compl. & Divers. & Title \\
\midrule
\multicolumn{9}{l}{\textit{Baselines}} \\
BM25 & \underline{0.56$_{\scriptstyle \pm 0.02}$} & \underline{0.87$_{\scriptstyle \pm 0.02}$} & \underline{0.84$_{\scriptstyle \pm 0.02}$} & \underline{0.93$_{\scriptstyle \pm 0.02}$} & 0.92$_{\scriptstyle \pm 0.02}$ & \underline{0.75$_{\scriptstyle \pm 0.02}$} & 1.06$_{\scriptstyle \pm 0.02}$ & \underline{0.54$_{\scriptstyle \pm 0.02}$} \\
Dense (MiniLM) & 0.39$_{\scriptstyle \pm 0.01}$ & 0.67$_{\scriptstyle \pm 0.01}$ & 0.65$_{\scriptstyle \pm 0.01}$ & 0.73$_{\scriptstyle \pm 0.02}$ & 0.78$_{\scriptstyle \pm 0.01}$ & 0.61$_{\scriptstyle \pm 0.02}$ & 0.94$_{\scriptstyle \pm 0.02}$ & 0.37$_{\scriptstyle \pm 0.01}$ \\
Hybrid ($\alpha$=0.5) & 0.49$_{\scriptstyle \pm 0.01}$ & 0.78$_{\scriptstyle \pm 0.01}$ & 0.76$_{\scriptstyle \pm 0.02}$ & 0.82$_{\scriptstyle \pm 0.02}$ & \underline{0.92$_{\scriptstyle \pm 0.01}$} & 0.74$_{\scriptstyle \pm 0.02}$ & \underline{1.06$_{\scriptstyle \pm 0.02}$} & 0.46$_{\scriptstyle \pm 0.01}$ \\
\midrule
Generative Retrieval & \textbf{0.71$_{\scriptstyle \pm 0.02}$}\textsuperscript{***} & \textbf{0.99$_{\scriptstyle \pm 0.01}$}\textsuperscript{***} & \textbf{1.04$_{\scriptstyle \pm 0.01}$}\textsuperscript{***} & \textbf{1.05$_{\scriptstyle \pm 0.01}$}\textsuperscript{***} & \textbf{1.15$_{\scriptstyle \pm 0.01}$}\textsuperscript{***} & \textbf{1.28$_{\scriptstyle \pm 0.02}$}\textsuperscript{***} & \textbf{1.39$_{\scriptstyle \pm 0.01}$}\textsuperscript{***} & \textbf{0.66$_{\scriptstyle \pm 0.02}$}\textsuperscript{***} \\
\bottomrule
\end{tabular}
}
\end{table*}

\paragraph{Hypothesis-to-Shelf Judge} This judge evaluates whether retrieved catalogue items successfully realise the generated shelf hypothesis. It conditions on the shelf title, subtitle, hypothesis, and rendered item list enriched with metadata such as genre, origin, year, and language. Table~\ref{tab:hypothesis-to-shelf-rubric} summarises the seven evaluated dimensions together with the overall judge score.

\begin{table*}[t]
\centering
\caption{Hypothesis-to-Shelf Judge rubric.}
\label{tab:hypothesis-to-shelf-rubric}
\scriptsize
\setlength{\tabcolsep}{4pt}
\renewcommand{\arraystretch}{1.08}
\begin{tabular}{@{}p{0.20\textwidth}p{0.765\textwidth}@{}}
\toprule
Dimension & What it measures \\
\midrule
Style Match &
Alignment between item-level style attributes and those specified by the
hypothesis, including genre, era, topic, format, and tone. \\
Item Relevance &
Whether each item is a credible representative of the described niche; filler,
popularity mismatches, and tangential items are penalised. \\
Shelf Coherence &
Whether the item set hangs together through era, scene, sound, or curatorial
logic, assessed independently of hypothesis fit. \\
Hypothesis Coverage &
Whether the shelf addresses the specific nuances named in the hypothesis; a
maximum score requires every distinct nuance to be reflected. \\
Completeness &
Whether canonical items associated with the hypothesis are present, capturing
whether the shelf would satisfy a knowledgeable user's baseline expectations. \\
Diversity &
Whether the shelf varies its items while remaining on-hypothesis; highly
concentrated shelves are penalised unless the concept explicitly requires it. \\
Title Promise Fulfilment &
Whether the items satisfy the explicit claims made by the title and subtitle; the
judge also records the main mismatch axis when that promise is unmet. \\
\bottomrule
\end{tabular}
\end{table*}

\subsection{Hypothesis quality}

We first evaluate whether the hypothesis generator produces personalised and sufficiently specific shelf hypotheses. 
Table~\ref{tab:hypotheses_generation_overall_scores} reports User-to-Hypothesis Judge results for the frontier-LLM generator used during model development. Generated hypotheses score strongly overall, with especially high hypothesis specificity and title quality, indicating that the system typically produces concrete and user-facing shelf concepts rather than vague recommendation prompts. Qualitative inspection of sampled outputs showed a similar pattern: most generated hypotheses appeared specific, plausible, and recognisably grounded in the intended user context.

\begin{table}[t]
\centering
\caption{Stage 1 hypothesis quality under the User-to-Hypothesis Judge (0--2 scale). Cells report mean score $\pm$ the 95\% bootstrap confidence-interval half-width from 10k resamples.}
\label{tab:hypotheses_generation_overall_scores}
\scriptsize
\setlength{\tabcolsep}{8pt}
\renewcommand{\arraystretch}{1.08}
\begin{tabular}{lc}
\toprule
\multicolumn{2}{c}{\textit{Quality by judged dimension}} \\
\midrule
Metric & Score \\
\midrule
Overall & 1.59$_{\scriptstyle \pm 0.01}$ \\
Taste Alignment & 1.59$_{\scriptstyle \pm 0.01}$ \\
Personalisation Depth & 1.45$_{\scriptstyle \pm 0.01}$ \\
Discovery Potential & 1.31$_{\scriptstyle \pm 0.01}$ \\
Hypothesis Specificity & 1.99$_{\scriptstyle \pm 0.00}$ \\
Title Quality & 1.76$_{\scriptstyle \pm 0.01}$ \\
\midrule
\multicolumn{2}{c}{\textit{Quality by content type}} \\
\midrule
Content Type & Avg. \\
\midrule
Album & 1.90$_{\scriptstyle \pm 0.01}$ \\
Artist & 1.98$_{\scriptstyle \pm 0.01}$ \\
Playlist & 1.86$_{\scriptstyle \pm 0.01}$ \\
Show & 0.66$_{\scriptstyle \pm 0.03}$ \\
Episode & 0.51$_{\scriptstyle \pm 0.05}$ \\
\bottomrule
\end{tabular}
\end{table} 
Because the production pipeline uses the distilled open-source generator described in Section~\ref{sec:pipeline}, we also ran a separate parity check on a fixed cohort of 800 production user profiles from a Stage~1 profile snapshot. The Frontier LLM baseline and distilled student were evaluated on the same 800 user IDs with the User-to-Hypothesis Judge. Their overall scores were effectively unchanged (78.3\% versus 78.2\%), with only small, mixed dimension-level differences. This supports using the distilled generator in production without making hypothesis generation the dominant source of quality loss, and motivates the stronger focus on catalogue fulfilment below.

Performance is strongest for music-oriented hypotheses, while spoken-word shelf concepts remain more challenging. Taken together, these findings suggest that the more consequential quality bottleneck lies downstream in catalogue fulfilment, which we evaluate next.

\subsection{Catalogue fulfilment quality}

We next evaluate whether Generative Retrieval fulfils shelf hypotheses more effectively than conventional text-retrieval approaches. To isolate the contribution of the Stage 2 fulfilment model, we compare Generative Retrieval against three retrieval baselines on the same 10,000 shelf hypotheses from the 1,000-user evaluation cohort: BM25~\cite{robertson1994okapi}, MiniLM~\cite{sentence-transformers-all-MiniLM-L12-v2} dense retrieval, and a linear equal-weight interpolation of BM25 and MiniLM scores. All methods fulfil within the hypothesis-specified content type and are evaluated using the same Hypothesis-to-Shelf Judge described above.

Table~\ref{tab:retrieval-all-dims} shows that Generative Retrieval outperforms all lexical and embedding-based fulfilment baselines across every judged dimension, with the largest gains in completeness, diversity, and hypothesis coverage. These improvements remain significant after Bonferroni correction across the 24 method-by-dimension comparisons considered in this analysis. The results support the central hypothesis of the paper: many shelf concepts require catalogue associations that are not recoverable through direct lexical or embedding similarity alone.

The relative performance gap was not uniform across content types. Lexical baselines were more competitive when catalogue entities contained rich curated descriptors that closely matched the language used in the shelf hypotheses, allowing direct text matching to recover a substantial fraction of the intended signal. BM25 particularly benefited in these settings. Performance degraded, however, when shelf concepts were expressed more indirectly through stylistic associations, broader cultural context, or combinations of attributes only weakly represented in item metadata.

In principle, some of this gap could be reduced through extensive manual descriptor engineering for each catalogue domain. However, maintaining such descriptors at production scale is expensive, content-type-specific, and difficult to align with the evolving language of generated shelf hypotheses. Generative Retrieval provides a scalable alternative by learning to map expressive shelf hypotheses directly to catalogue entities without requiring exhaustive manual descriptor design.

\subsection{Effect of candidate selection and shelf alignment}

We next evaluate whether candidate selection and shelf alignment improve the quality and coherence of the final displayed shelf. This analysis compares shelves immediately after catalogue fulfilment with shelves after the Stage 3 alignment process using the Hypothesis-to-Shelf Judge described above.

Table~\ref{tab:production_generation_overall_scores} isolates the contribution of the alignment stage by comparing fulfilled shelves before alignment against production shelves after alignment. The pre-alignment cohort contains 10,000 shelves, while the post-alignment production cohort contains over 16,000 shelves; the two sets contain no shared shelf identifiers. We therefore use two-sided Welch independent-samples $t$-tests with Bonferroni correction across the eight reported dimensions (\textsuperscript{***} denotes $p<0.001$ after correction).

\begin{figure}[t]
    \centering
    \includegraphics[width=\linewidth]{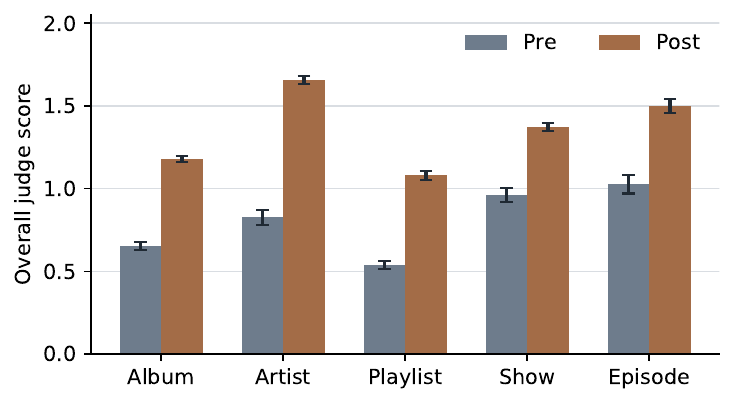}
    \caption{Hypothesis-to-Shelf Judge score by content type before and after Stage 3 shelf alignment, with 95\% bootstrap confidence intervals.}
    \Description{Chart comparing Hypothesis-to-Shelf Judge scores before and after shelf alignment across content types; post-alignment scores are higher for every content type.} \label{fig:hypothesis-to-shelf-by-content-type}
\end{figure}

Candidate selection and shelf alignment substantially improve overall shelf quality, increasing the overall judge score from 0.71 to 1.27 (+78\%). Improvements are observed across every judged dimension, with the largest gains in title-promise fulfilment (+99\%), shelf coherence (+56\%), item relevance (+52\%), and style match (+52\%). Figure~\ref{fig:hypothesis-to-shelf-by-content-type} shows that the same pattern holds across content types: all groups improve after alignment, with especially large relative gains for playlists and artists.

These findings support the motivation for Stage 3. Retrieval quality alone is insufficient for shelf recommendation because the shelf text acts as a semantic promise about the entire row. Candidate selection and shelf alignment substantially improve both the coherence of the retrieved set and the consistency between the shelf concept and the final catalogue entities shown to users.

\begin{table}[t]
\centering
\caption{Effect of Stage 3 shelf alignment under the Hypothesis-to-Shelf Judge (0--2 scale). Cells report mean score $\pm$ the 95\% bootstrap confidence-interval half-width from 10k resamples.}
\label{tab:production_generation_overall_scores}
\scriptsize
\begin{tabular}{lccc}
\toprule
Metric & Pre & Post & $\Delta$ (\%) \\
\midrule
Overall & 0.71$_{\scriptstyle \pm 0.02}$ & \textbf{1.27$_{\scriptstyle \pm 0.01}$} & +78\%\textsuperscript{***} \\
Style Match & 0.99$_{\scriptstyle \pm 0.01}$ & \textbf{1.51$_{\scriptstyle \pm 0.01}$} & +52\%\textsuperscript{***} \\
Item Relevance & 1.04$_{\scriptstyle \pm 0.01}$ & \textbf{1.58$_{\scriptstyle \pm 0.01}$} & +52\%\textsuperscript{***} \\
Shelf Coherence & 1.05$_{\scriptstyle \pm 0.01}$ & \textbf{1.64$_{\scriptstyle \pm 0.01}$} & +56\%\textsuperscript{***} \\
Hypothesis Coverage & 1.15$_{\scriptstyle \pm 0.01}$ & \textbf{1.37$_{\scriptstyle \pm 0.01}$} & +19\%\textsuperscript{***} \\
Completeness & 1.28$_{\scriptstyle \pm 0.02}$ & \textbf{1.46$_{\scriptstyle \pm 0.01}$} & +14\%\textsuperscript{***} \\
Diversity & 1.39$_{\scriptstyle \pm 0.01}$ & \textbf{1.74$_{\scriptstyle \pm 0.01}$} & +25\%\textsuperscript{***} \\
Title Promise & 0.66$_{\scriptstyle \pm 0.02}$ & \textbf{1.31$_{\scriptstyle \pm 0.01}$} & +99\%\textsuperscript{***} \\
\bottomrule
\end{tabular}
\end{table}
 
\subsection{Online performance under randomised exposure}

We finally evaluate whether hypothesis-driven shelves perform competitively under uniform random exposure on Home. Unlike the earlier stages, which evaluate intermediate artefacts and semantic quality, this stage focuses on observed user behaviour when generated shelves are served on Home.

Shelf engagement on Home is strongly influenced by ranking position and by how the production ranker selects and orders shelves, making direct comparison between shelf families difficult under the standard serving policy. To reduce these effects, we use a standard randomised-exposure protocol for industrial recommender evaluation~\cite{randomisedExposureRecsys}. A small fraction of Home requests are assigned to uniform random exploration, where shelf ordering is randomised independently of the production ranker. Assignment occurs at the request level rather than the user level, allowing exploration impressions to be collected without persistent exposure to randomised feeds. Under this protocol, we compare hypothesis-driven shelves against existing production shelf families under uniform random exposure.

Table~\ref{tab:online_shuffle_pool_summary} summarises the content-type-specific shuffle-pool comparisons. Hypothesis-driven shelves have the strongest observed mean in the album pool (+36\%) and episode pool (+2\%), and rank second in the artist and show pools. Performance is weaker in the playlist and show pools, where established production shelves retain an advantage. The largest gap appears in the show pool ($-41\%$), indicating clear headroom for improving podcast shelf generation. Although performance varies across content types, hypothesis-driven shelves are competitive with strong existing shelves in the majority of pools, while substantially expanding the flexibility and diversity of personalised recommendation supply beyond what a fixed template inventory can cover.

\begin{table}[t]
\centering
\caption{Uniform-random-exploration summary across content-type shuffle pools. For each pool, we report the strongest hypothesis-driven shelf and the strongest classic comparison shelf observed in that pool. Cells show 30-second stream rate in percentage points as mean $\pm$ the 95\% bootstrap confidence-interval half-width; $\Delta$ is the relative percent difference between the hypothesis-driven shelf and the classic comparator. These are descriptive within-pool comparisons, not pooled causal effect estimates.}
\label{tab:online_shuffle_pool_summary}
\scriptsize
\begin{tabular}{lcccc}
\toprule
Content type & Hypothesis-driven (\%) & Best classic (\%) & Rank & $\Delta$ (\%) \\
\midrule
Album & 1.20$_{\scriptstyle \pm 0.13}$ & 0.88$_{\scriptstyle \pm 0.09}$ & 1 / 10 & +36\% \\
Artist & 0.82$_{\scriptstyle \pm 0.17}$ & 0.89$_{\scriptstyle \pm 0.06}$ & 2 / 9 & -8\% \\
Playlist & 0.92$_{\scriptstyle \pm 0.09}$ & 1.07$_{\scriptstyle \pm 0.07}$ & 5 / 15 & -14\% \\
Show & 0.92$_{\scriptstyle \pm 0.31}$ & 1.57$_{\scriptstyle \pm 0.11}$ & 2 / 6 & -41\% \\
Episode & 0.63$_{\scriptstyle \pm 0.20}$ & 0.62$_{\scriptstyle \pm 0.08}$ & 1 / 7 & +2\% \\
\bottomrule
\end{tabular}
\end{table}
 
\section{Conclusion}
\label{sec:conclusion}

We introduced a hypothesis-driven framework for generating personalised Home shelves in a production setting. The system uses explicit shelf hypotheses as intermediate planning representations, separating personalised shelf planning from catalogue fulfilment before refining retrieved candidates through shelf alignment and serving them within the existing Home ranking system.

This decomposition yields a recommendation architecture that is both expressive and operationally scalable. Hypothesis generation remains strong after distillation into compact production models, Generative Retrieval substantially improves catalogue fulfilment over lexical and embedding-based baselines, and candidate selection and shelf alignment significantly improve shelf coherence and title-promise fulfilment. Under randomised exposure on Home, the resulting shelves achieve engagement that varies by content type and is competitive with strong production alternatives in some settings while substantially expanding the flexibility and diversity of personalised recommendation supply.

More broadly, our results suggest that recommendation interfaces can be generated through explicit shelf concepts rather than fixed template inventories, enabling personalised recommendation surfaces that scale beyond hand-authored shelf taxonomies.

Several directions remain open for future work, including evaluation under standard production ranking. One direction is tighter integration between planning and fulfilment, allowing retrieval to adapt more directly to user-specific context while preserving controllability and evaluation boundaries. Another is moving beyond fully precomputed serving toward near-real-time shelf generation, enabling faster adaptation to evolving user behaviour while retaining production-grade reliability and evaluation guarantees.

\begin{acks}
  We thank Sanjana Kacholia for her helpful feedback and suggestions on the camera-ready manuscript.

  \paragraph{Use of generative AI}
  Parts of this work were developed with the assistance of generative AI tools, which served as aids for writing data-processing and visualization code, as well as for refining drafts of the text. All generated material was reviewed, verified, and edited by the authors, who take full responsibility for the final content.
\end{acks}

\bibliographystyle{ACM-Reference-Format}
\bibliography{references}
\end{document}